\documentclass[11pt]{article}
\usepackage{amssymb}
\usepackage{epsfig}

\newcommand{\BE}{\begin{equation}}
\newcommand{\EE}{\end{equation}}
\newcommand{\BA}{\begin{eqnarray}}
\newcommand{\EA}{\end{eqnarray}}

\begin{document}

\begin{titlepage}
\begin{center}

   {\LARGE{\bf An easy reading of modern ether-drift experiments}}

\vspace*{14mm} {\Large  M. Consoli and E. Costanzo}
\vspace*{4mm}\\
{
Istituto Nazionale di Fisica Nucleare, Sezione di Catania \\
Dipartimento di Fisica e Astronomia dell' Universit\`a di Catania \\
Via Santa Sofia 64, 95123 Catania, Italy \\ }
\end{center}
\begin{center}
{\bf Abstract}
\end{center}
Modern ether-drift experiments look for a preferred reference frame
searching for modulations of the beat note of two optical resonators
that might be induced by the Earth's rotation. We present a compact
formalism to evaluate the signal for most experiments where two
arbitrary gaseous media fill the resonating cavities. Our
predictions can provide useful hints to optimize the experimental
set up and the data taking.

\end{titlepage}

{\bf 1}.~In modern ether-drift experiments with optical resonators,
the search for the possible existence of a preferred reference frame
is performed by looking for modulations of the signal that might be
induced by the Earth's rotation. Descriptions of this important
effect are already available in the literature. For instance, within
the SME model \cite{sme} the relevant formulas are given in the
appendix of Ref.\cite{mewes} and for the RMS test theory \cite{rms}
one can look at Ref.\cite{applied}. However, either due to the great
number of free parameters (19 in the SME model) and/or to the
restriction to a definite experimental set up, it is not always easy
to adapt these papers to the various experimental conditions. For
this reason, in this Letter, we will present a set of compact
formulas that can be immediately used by the reader to evaluate the
signal when two arbitrary gaseous media fill the resonating
cavities. The formalism covers most experimental set up including
the very recent type of experiment proposed in Ref.\cite{luiten}.

In our presentation one clearly understands that the Earth's
rotation enters only through two quantities, $v=v(t)$ and
$\theta_0=\theta_0(t)$, respectively the magnitude and the angle
associated with the projection of the unknown cosmic Earth's
velocity ${\bf{V}}$ in the plane of the interferometer. At the same
time, our predictions can provide useful hints to optimize the
experimental set up and the data taking.

\vskip 10 pt

{\bf 2.}~Once the angle $\theta_0$ is conventionally defined when
one of the arms of the interferometer is oriented to the North point
in the laboratory (counting $\theta_0$ from North through East), we
can immediately use the formulas given by Nassau and Morse
\cite{nassau}. These are valid for short-term observations, say 3-4
days, where there are no appreciable changes in the cosmic velocity
due to changes in the Earth's orbital velocity around the Sun so
that the only time dependence is due to the Earth's rotation.

In this approximation, introducing the magnitude $V$ of the full
Earth's velocity with respect to a hypothetic preferred frame
$\Sigma$, its right ascension $\alpha$ and angular declination
$\gamma$, we get

\BE
       \cos z(t)= \sin\gamma\sin \phi + \cos\gamma
       \cos\phi \cos(\tau-\alpha)
\EE \BE
       \sin z(t)\cos\theta_0(t)= \sin\gamma\cos \phi -\cos\gamma
       \sin\phi \cos(\tau-\alpha)
\EE \BE
       \sin z(t)\sin\theta_0(t)= \cos\gamma\sin(\tau-\alpha) \EE
\BE \label{projection}
       v(t)=V \sin z(t) ,
\EE
where $z=z(t)$ is the zenithal distance of ${\bf{V}}$, $\phi$ is the
latitude of the laboratory and $\tau=\omega_{\rm sid}t$ is the
sidereal time of the observation in degrees ($\omega_{\rm sid}\sim
{{2\pi}\over{23^{h}56'}}$).

Let us now consider two orthogonal cavities oriented for simplicity
to North (cavity 1)  and East (cavity 2) in the laboratory frame.
They are filled with two different gaseous media with refractive
indices ${\cal N}_i$ (i=1,2) such that ${\cal N}_i=1+\epsilon_i$,
and $0\leq \epsilon_i \ll 1$. The frequency in each cavity is \BE
\nu_i(\theta_i)=\bar{u}'_i(\theta_i)k_i \EE  and the frequency shift
is \BE \Delta\nu=\nu_1(\theta_1)-\nu_2(\theta_2) \EE In the above
relations we have introduced the parameters $k_i$ \BE
k_i={{m_i}\over{2L_i}}\EE where $m_i$ are integers fixing the cavity
modes, $L_i$ are the cavity lengths and $\bar{u}'_i(\theta_i)$
denote the two-way speeds of light, as measured in the Earth's
frame, $\theta_i$ being the angle between ${\bf{V}}$ and the axis of
the i-th cavity.

Following the point of view of Refs.\cite{pla,guerra,reply}, that no
observable Fresnel's drag has ever been detected in the gaseous
regime, we shall assume that the two speeds of light ${{c}\over{
{\cal N}_i}}$ are seen isotropic in the preferred frame $\Sigma$.
Using Lorentz transformations to connect to the Earth's frame, one
then obtains to ${\cal O}(V^2/c^2)$
 \cite{pla} \BE \label{twoway}
\bar{u}'_i(\theta)={{c}\over{ {\cal N}_i
}}[1-(A_i+B_i\sin^2\theta){{V^2}\over{c^2}}] \EE with \BE
\label{lorentz} A_i={{{\cal N}^2_i-1}\over{ {\cal
N}^2_i}}~~~~~~~~~~~~B_i=-{{3}\over{2}}A_i \EE We emphasize that the
structure in Eq.(\ref{twoway}), although obtained in connection with
Eqs.(\ref{lorentz}) by using Lorentz transformations, remains also
valid under the more general assumptions of the RMS test theory
\cite{rms}. As such, if $A_i$ and $B_i$ are considered as free
parameters, it provides a physical framework that is equivalent to
the RMS model.

Introducing the unit vectors $\hat{u}_i$ fixing the direction of the
two cavities and the projection ${\bf{v}}$ of the full ${\bf{V}}$ in
the interferometer's plane one finds  \BE
V^2\sin^2\theta_i=V^2(1-\cos^2\theta_i)=V^2-(\hat{u}_i\cdot {\bf{v}}
)^2 \EE so that ($v=|{\bf{v}}|$) \BE V^2\sin^2\theta_1=V^2-
v^2\cos^2\theta_0 \EE and \BE V^2\sin^2\theta_2=V^2-
v^2\sin^2\theta_0\EE Therefore, defining the reference frequency
$\nu_0={{c k_1}\over{{\cal N}_1}}$ and introducing the parameter
$\xi$ through \BE \xi={{ {\cal N}_1 k_2}\over{{\cal N}_2 k_1 }} \EE
one finds the relative frequency shift \BE \label{general} {{\Delta
\nu(t)}\over{\nu_0}}=1- \xi + {{V^2}\over{c^2}}[\xi(A_2 +B_2)
-(A_1+B_1)] + {{v^2(t)}\over{c^2}}[B_1\cos^2\theta_0(t) - \xi
B_2\sin^2\theta_0(t)] \EE  For a symmetric apparatus where ${\cal
N}_1={\cal N}_2$, $A_1=A_2$, $B_1=B_2=B$ and $\xi=1$, one finds \BE
\label{symm}{{\Delta \nu(t)_{\rm symm}}\over{\nu_0}} = B
{{v^2(t)}\over{c^2}} \cos2\theta_0(t) \EE On the other hand for a
non-symmetric apparatus of the type considered in Ref.\cite{luiten}
with $L_1=L_2=L$, but where one can conveniently arrange ${\cal
N}_1=1$ (up to negligible terms) so that $A_1\sim B_1 \sim 0$,
denoting ${\cal N}_2={\cal N}$, $A_2=A$, $B_2=B$,
${{m_2}\over{m_1}}={\cal P}$, we find \BE \label{asymm} {{\Delta
\nu(t)}\over{\nu_0}}=1- {{{\cal P}\over{\cal N}}}+ {{{\cal
P}\over{\cal N}}}{{V^2}\over{c^2}}(A +B) - B{{{\cal P}\over{\cal
N}}}{{v^2(t)}\over{c^2}} \sin^2\theta_0(t) \EE To consider
experiments where one or both resonators are placed in a state of
active rotation (at a frequency $\omega_{\rm rot} \gg \omega_{\rm
sid}$), it is convenient to modify Eq.(\ref{general}) by rotating
the resonator 1 by an angle $\delta_1$ and the resonator 2 by an
angle $\delta_2$ so that the last term in Eq.(\ref{general}) becomes
\BE {{v^2(t)}\over{ c^2}}[B_1\cos^2(\delta_1-\theta_0(t)) - \xi
B_2\sin^2(\delta_2-\theta_0(t))] \EE Therefore, in a fully symmetric
apparatus where ${\cal N}_1={\cal N}_2$, $A_1=A_2$, $B_1=B_2=B$ and
$\xi=1$ and both resonators rotate, as in Ref.\cite{schiller},
setting \BE\delta_1=\delta_2=\omega_{\rm rot}t \EE one obtains \BE
\label{symm2}  {{\Delta \nu(t)_{\rm symm}}\over{\nu_0}}= B
{{v^2(t)}\over{c^2}} \cos2( \omega_{\rm rot}t -\theta_0(t)) \EE On
the other hand, if only one resonator rotates, as in
Ref.\cite{peters}, setting $\delta_1=0$ and $\delta_2=\omega_{\rm
rot}t$ one obtains the alternative result \BE \label{asymm2}
{{\Delta \nu(t)}\over{\nu_0}}= B {{v^2(t)}\over{2
c^2}}[\cos2\theta_0(t) + \cos2( \omega_{\rm rot}t -\theta_0(t))] \EE
By first filtering the signal at the frequency $\omega=\omega_{\rm
rot} \gg \omega_{\rm sid}$, the main difference between the two
expressions is an overall factor of two.\vskip 10 pt

{\bf 3.}~Let us now return to the general case of a non-rotating set
up Eq.(\ref{general}). Using Eqs.(1-4) we obtain the simple Fourier
expansion \BE {{\Delta \nu(t)}\over{\nu_0}}=1-\xi + (f_0+f_1\sin\tau
+f_2\cos\tau +f_3\sin 2\tau+ f_4\cos2\tau )\EE where \BE
f_0={{V^2}\over{c^2}}[ \xi(A_2 +B_2) -(A_1+B_1) +
B_1(\sin^2\gamma\cos^2\phi+ {{1}\over{2}}\cos^2\gamma\sin^2\phi)
-{{1}\over{2}}\xi B_2 \cos^2\gamma ]\EE \BE \label{f12}
f_1=-{{1}\over{2}}{{V^2}\over{c^2}}B_1\sin 2\gamma\sin 2\phi \sin
\alpha
~~~~~~~~~~~~~~~~~~~~~~~f_2=-{{1}\over{2}}{{V^2}\over{c^2}}B_1\sin
2\gamma\sin 2\phi \cos \alpha \EE \BE \label{f34}
f_3={{1}\over{2}}{{V^2}\over{c^2}}(B_1\sin^2\phi +\xi
B_2)\cos^2\gamma\sin 2\alpha
~~~~~~~~~~f_4={{1}\over{2}}{{V^2}\over{c^2}}(B_1\sin^2\phi +\xi
B_2)\cos^2\gamma\cos 2\alpha~~\EE Since the mean signal is most
likely affected by systematic effects, one usually concentrates on
the daily modulation. In this case, assuming that $f_1$, $f_2$,
$f_3$ and $f_4$ can be extracted to good accuracy from the
experimental data, one can try to obtain a pair of angular variables
through the two independent determinations of $\alpha$ \BE
\label{alpha} \tan \alpha= {{f_1}\over{f_2}}~~~~~~~~~~~~~~~\tan
2\alpha= {{f_3}\over{f_4}}\EE and the relation \BE \tan |\gamma|
={{|B_1\sin^2\phi +\xi B_2|}\over{|2 B_1\sin 2\phi|}}~ \sqrt{
{{f^2_1+f^2_2}\over{f^2_3+f^2_4}}} \EE Notice that, since the
ether-drift is a 2nd-harmonic effect, the pair $(\alpha,\gamma)$
cannot be distinguished from the pair $(\alpha+\pi,-\gamma)$. Notice
also that two dynamical models that predict the same anisotropy
parameters up to an overall re-scaling $B_i \to \lambda B_i$ would
produce the same $|\gamma|$ from the experimental data.

Finally for a symmetric apparatus, where $B_1=B_2=B$ and $\xi=1$,
one obtains the simpler relation \BE \label{gamma} \tan |\gamma|
={{1+\sin^2\phi }\over{|2 \sin 2\phi|}}~ \sqrt{
{{f^2_1+f^2_2}\over{f^2_3+f^2_4}}} \EE where any reference to the
anisotropy parameters drops out.  \vskip 10 pt

{\bf 4.}~Summarizing: starting from the hypothetical observation of
a non-trivial daily modulation of the signal in some ether-drift
experiment, one might meaningfully consider the possibility of a
preferred reference frame. For instance, for a symmetric apparatus
one could try to extract from the data the product
$K=B{{V^2}\over{c^2}}$ and, using Eqs.(\ref{alpha}) and
(\ref{gamma}), a pair of angular values $(\alpha,\gamma)$. Of
course, in this case, by suitably changing the gaseous medium within
the cavities, one should also try to check the trend predicted in
Eq.(\ref{lorentz}), namely \BE {{K'}\over{K''}}\sim {{ {\cal
N}'-1}\over{ {\cal N}''-1 }}\EE On the other hand, for a
non-symmetric apparatus of the type proposed in Ref.\cite{luiten},
where one can conveniently fix the cavity oriented to North to have
$N_1=1$ (up to negligible terms), by using Eqs.(\ref{lorentz}) one
would predict $B_1\sim 0$ in Eqs.(\ref{f12}) and (\ref{f34}) so that
all time dependence should be due to $B_2$. Thus the modulation of
the signal should be a pure $\omega=2\omega_{\rm sid}$ effect with
no appreciable contribution at $\omega=\omega_{\rm sid}$. This is
another sharp prediction that should be preliminarily checked.

For a deeper analysis, it is important to recall that the
ether-drift, if it exists, is a 2nd-harmonic effect. Therefore, in a
single session, the direction $(\alpha,\gamma)$ cannot be
distinguished from the opposite direction $(\alpha+\pi,-\gamma)$.
For this reason, a whole set j=1,2..M of short-term experimental
sessions should be performed in different periods along the Earth's
orbit to obtain an overall consistency check.

Notice that for a complete description of the observations over a
one-year period, it is not necessary to modify the simple formulas
Eqs.(\ref{f12}) and (\ref{f34}) and introduce explicitly the further
modulations associated with the orbital frequency $\Omega_{\rm
orb}\sim {{2\pi}\over{1 ~{\rm year}}}$. Rather, by plotting on the
celestial sphere all directions defined by the various
$(\alpha_j,\gamma_j)$ pairs obtained in the various short-term
observations one can try to reconstruct the Earth's ``aberration
circle''. If this will show up, one can determine the mean magnitude
of the cosmic velocity $\langle V\rangle $ from the angular opening
of the circle and from the known value of the orbital Earth's
velocity $\sim 30 $ km/s. In this way, given the value of $\langle K
\rangle$, one will be able to disentangle $\langle V\rangle $ from
$B$ and get a definitive test of models that predict the absolute
magnitude of the anisotropy parameter.

\vfill\eject


\begin{thebibliography} {99}
\bibitem{sme}
D. Colladay and V. A. Kostelecky, Phys. Rev. {\bf D55} (1997) 6760;
{\bf 58} (1998) 116002; R. Bluhm, et al., Phys. Rev. Lett. {\bf 88}
(2002) 090801.
\bibitem{mewes}
V. A. Kostelecky and M. Mewes, Phys. Rev. {\bf D66} (2002) 056005.
\bibitem{rms}
H. P. Robertson, Rev. Mod. Phys. {\bf 21}, 378 (1949); R. M.
Mansouri and R. U. Sexl, Gen. Rel. Grav. {\bf 8}, 497 (1977).
\bibitem{applied}
H. M\"uller, et al. Appl. Phys. {\bf B77} (2003) 719.
\bibitem{luiten}
S. Dawkins and A. Luiten, "Testing the standard model of physics",
Presentation at the Australian Institute of Physics 17th National
Congress 2006, Brisbane December 2006.
\bibitem{nassau}
J. J. Nassau and P. M. Morse, see Astrophys. Journ. {\bf 65} (1927)
73.
\bibitem{pla}
M. Consoli and E. Costanzo, Phys. Lett. {\bf A333} (2004) 355; N.
Cim. {\bf 119B} (2004) 393 [arXiv:gr-qc/0406065].
\bibitem{guerra}
V. Guerra and R. De Abreu, Phys. Lett. {\bf A361} (2007) 509.
\bibitem{reply}
M. Consoli and E. Costanzo, Phys. Lett. {\bf A361} (2007) 513.
\bibitem{schiller}
P. Antonini, et al., Phys. Rev. {\bf A71}, 050101(R)(2005)
[arXiv:gr-qc/0504109].
\bibitem{peters}
S. Herrmann, et al., Phys. Rev. Lett. {\bf 95}, 150401 (2005)
[arXiv:physics/0508097].







\end{thebibliography}
\end{document}